# Towards a Feminist Metaethics of AI

Anastasia Siapka
Centre for IT & IP Law
Katholieke Universiteit Leuven
Leuven, Belgium
anastasia.siapka@kuleuven.be

## ABSTRACT

The proliferation of Artificial Intelligence (AI) has sparked an overwhelming number of AI ethics guidelines, boards and codes of conduct. These outputs primarily analyse competing theories, principles and values for AI development and deployment. However, as a series of recent problematic incidents about AI ethics/ethicists demonstrate, this orientation is insufficient. Before proceeding to evaluate other professions, AI ethicists should critically evaluate their own; yet, such an evaluation should be more explicitly and systematically undertaken in the literature. I argue that these insufficiencies could be mitigated by developing a research agenda for a feminist metaethics of AI. Contrary to traditional metaethics, which reflects on the nature of morality and moral judgements in a non-normative way, feminist metaethics expands its scope to ask not only what ethics *is* but also what our engagement with it *should be* like. Applying this perspective to the context of AI, I suggest that a feminist metaethics of AI would examine: (i) the continuity between theory and action in AI ethics; (ii) the real-life effects of AI ethics; (iii) the role and profile of those involved in AI ethics; and (iv) the effects of AI on power relations through methods that pay attention to context, emotions and narrative.

## CCS CONCEPTS

• Computing methodologies~Artificial intelligence • Social and professional topics~Computing / technology policy

## KEYWORDS

Artificial Intelligence, AI ethics, feminist philosophy, metaethics, ethics-washing

## 1 Does AI Ethics Have a Problem?

Over the past years, the proliferation of Artificial Intelligence (AI) algorithms in a multitude of, often high-stake, contexts has sparked intense and sustained debate about the risks of AI and, as a result, its appropriate governance. As part of this debate, an overwhelming number of AI ethics guidelines [33, 81], boards [68, 46], codes of conduct [2] and even dedicated journals [84] have emerged across the globe, with academics and practitioners in law, philosophy, and computer science as well as policymakers and industry players all claiming expertise in the field of so-called AI ethics [30]. When it comes to the outputs of this field, these consist primarily in analyses of competing theories, principles, and values that should guide the development and deployment of AI. Secondarily, a limited number of meta-analyses assemble different ethics resources and group them into inventories and databases [3, 64, 86] or, moving a step further, compare them to identify common themes [22, 69, 87, 37].

Both these types of outputs admittedly bring value to the AI ethics debate. Nonetheless, the adequacy of this debate, as is currently performed, along with that of the people partaking in it are called into question by a series of recent incidents. In a 2019 op-ed, Metzinger [53], a member of the European Commission's High-Level Expert Group on AI, known as AI HLEG, accused the group itself of 'ethics washing made in Europe' and warned of 'fake ethics'. It was the same year that Google terminated its AI ethics board only a week after its formation due to backlash over the inclusion of an openly anti-LGBT member [76], whereas Dr Joichi Ito, former director of the MIT Media Lab and AI ethics expert, was found to have financial ties with Jeffrey Epstein [66]. Still in 2019, Facebook funded an AI ethics research centre at the Technical University of Munich, igniting concerns about undue influence on academic research and conflict of interest [85]. Moving on to 2020, the international human rights organisation Access Now resigned from the Partnership on AI (PAI), a coalition initially founded by Amazon, Facebook, Google, IBM and Microsoft, stating that they 'did not find that PAI influenced or changed the attitude of member companies or encouraged them to respond to or consult with civil society on a systematic basis' [38]. Near the end of the same year, it was revealed that Google was asking its researchers to 'strike a positive tone' in their outputs [14], while the sudden exit of AI ethicist Dr Timnit Gebru from the company, due to a paper highlighting the risks of Google's machine-learning language models, alongside the firing of AI ethicist Dr Margaret Mitchell a few months later garnered widespread media attention [62].

Incidents such as the above have a negative impact on AI ethics itself as an intellectual enterprise, since they undermine its legitimacy and contribution. Equally, they might harm the actual people and societies affected by AI and its governance. Hence, they constitute a cause for concern and invite second-order reflection on what AI ethics ought to be, how it should work, who should be doing it and who should be paying for it. Yet, in the relevant literature, such reflection is neither prioritised nor performed in a systematic manner.

By way of illustration, in their survey of the relevant literature, Gordon and Nyholm [30] identify the following topics among the main debates in AI ethics: machine ethics; autonomous systems; machine bias; the problem of opacity; machine consciousness; the moral status of AI; singularity and value alignment; and a category of other, various issues. Along the same lines, Müller's [60] review of the main debates in AI ethics includes: privacy and surveillance; manipulation of behaviour; the opacity of AI systems; bias in decision systems; human-robot interaction; automation and employment; autonomous systems; machine ethics; artificial moral agents; and the singularity. Both these reviews acknowledge critical concerns about the state of the AI ethics field, but such concerns are not listed among the field's main debates and so are relegated to only brief remarks. Similarly, although scholars have independently voiced critiques of the ways in which AI ethics is pursued—critiques that are also included in the remainder of this paper—these have been relatively under-theorised, especially by comparison with the AI ethics outputs outlined at the beginning of the paper.

Moreover, in spite of the fact that the diversity of existing criticism is certainly applauded, we need a way of systematically speaking about the state of AI ethics as a field. Although much of the literature focuses on identifying a suitable ethical framework for the development and deployment of AI, there have not been comparable efforts for the identification or construction of frameworks within which to carry out a second-order reflection on AI ethics. This omission is not negligible, since such a reflection, if organised on the basis of a specific theoretical framework, would benefit from a blueprint to guide and structure the discussion, methods and concepts upon which the discussion could draw as well as standards by which its success could be evaluated.

In this paper, I set out to argue that the foregoing insufficiencies could be mitigated by developing a new research agenda for a feminist metaethics of AI. To that end, in a first instance, I explain what feminist metaethics entails in general and, in a second instance, I elaborate on its application to the context of AI.

## 2 What Does Feminist Metaethics Mean?

Ethics, the philosophical branch devoted to the study of goodness and morality, is traditionally divided into three sub-disciplines: normative ethics, applied ethics and metaethics. Simply put, normative ethics asks what makes an action/outcome right or wrong or a person good or bad (Question 1). The answers to Question 1 can, in turn, be applied to actual and usually controversial cases or domains. So, applied ethics asks if a specific action, e.g. abortion or euthanasia, is right or wrong or how normative theories should be applied to a specific domain or practice, e.g. medicine (Question 2). Both Questions 1 and 2 are considered first-order questions. Metaethics, though, examines how humans reach moral judgements in response to Questions 1 and 2, and so moves to a level that is more abstract than both normative and applied ethics. For Schroeder [72:674], 'metaethics concerns questions *about* normative inquiry, rather than questions *within* normative inquiry'. It asks, by way of illustration, if moral judgements can be objectively true or false or if they represent mere opinions or desires (Question 3). In that sense, Question 3 encompasses second-order questions, i.e. questions about the status of first-order ethical questions.

In trying to describe and explain the nature of actual morality and moral judgements, such second-order questions are traditionally non-normative. To approach these, philosophers have defended competing theories, drawing mainly upon the sub-fields of metaphysics, epistemology, phenomenology, moral psychology, philosophy of mind and philosophy of language, thereby rendering metaethics a highly theoretical activity. Therefore, metaethics 'has nothing to say concerning the rightness of acts, makes no moral recommendations whatsoever, is entirely neutral as to which moral principles one is to adopt to guide one's affairs' [77:95].

Specifically, following Fisher [20], the most common divisions of theories within metaethics are: realism and non-realism; cognitivism and non-cognitivism; naturalism and non-naturalism; and, lastly, internalism and externalism. According to realists, 'moral properties exist and are in some way independent from people's judgements', whereas non-realists deny the existence of moral properties or facts. For cognitivists, moral claims express the speaker's beliefs, which can be true or false, while for non-cognitivists, moral claims express non-belief states, such as the speaker's emotions, and so cannot be true or false. When it comes to naturalists and non-naturalists, their difference lies in that the former accept only 'those things that would appear in the scientific picture of what exists' as existing, whereas the latter take a broader view. Finally, internalism and externalism refer to one's motivation: contrary to externalists, internalists maintain that there is a necessary connection between the moral judgements we make and our behaviour in keeping with such judgements.

On the whole, defined as 'the attempt to understand the metaphysical, epistemological, semantic, and psychological, presuppositions and commitments of moral thought, talk, and practice' [71], metaethics is a descriptive project that not only refrains from evaluating extant morality but can even be viewed as maintaining and legitimising it from a place of authority. This non-normative understanding of metaethics is questioned by feminist philosophy. Although feminist philosophy 'begins with attention to women, to their roles and locations' in the sense of focusing on moral and political topics related to women's oppression, its distinct contribution to philosophy is not confined to this selection of topics but likewise extends to critical discussions of 'mainstream philosophical views and methods' [51].

Given the origins of feminist philosophy in a movement for political change and its concomitant practical orientation towards understanding and ending women's oppression, its proponents 'generally share [...] a commitment to normativity and social change; they are never content to analyze things just as they are but are instead looking for ways to overcome sexist practices and institutions' [51]. They acknowledge that there is no such thing as an ideal observer; rather, a theorist's assumptions and choice of topics are influenced by social position and group membership. Accordingly, they argue that philosophical theorising is dominated by white, male, upper-middle-class scholars and is thereby likely to both 'reflect the interests and concerns of a small, particularly privileged, subset of the population' and 'overlook the effects that social systems of power have on everyone' [73:308]. Thus, feminist philosophers have been suspicious of traditional moral theories and have articulated critical questions about these, such as 'who makes them? for whom do they make them? about what or whom are the theories? why? how are theories tested? what are the criteria for such tests and where did the criteria come from?' as well as '[w]hy do we engage in *this* activity and what effect do we think it ought to have?' [45:578-579].

Seen through this reflexive perspective of feminist philosophy, '[m]eta-ethics doesn't have to be technical, non-moral and far removed from real life. We can also take metaethics to include asking distinctly ethical questions about doing and teaching ethics' [40]. Hence, metaethics could open up its scope to include a descriptive/non-normative project together with a normative one, where the latter may complement rather than replace the former. Subsequently, metaethics would also be asking how we should (or should not) be doing ethics in lieu of merely asking how we are doing ethics [40]. The scholarship that deals with these questions is not necessarily new but has been recently integrated and systematised by philosophers such as Superson [78], Srinivasan [74] and Jongepier [40] to create the sub-field of 'feminist metaethics'.[1]

Other moral theorists have in the past defended a rejection of the dichotomy between first- and second-order moral claims. Most notably, Dworkin [16:88] argues against 'archimedean' theories that 'purport to stand outside a whole body of belief, and to judge it as a whole from premises or attitudes that owe nothing to it'. As a result, he considers all second-order moral claims to actually be first-order ones, and thereby normative, and views them both as inextricably linked [17]. Similarly, Gewirth [29] casts doubt on the alleged independence of metaethics from normative ethics and seeks to demonstrate that metaethics is normative.

Olafson [67] and Taylor [80] also unveil normative dimensions of metaethics. Nonetheless, this rejection of the first- and second-order dichotomy in moral claims does not suffice on its own for a research programme to count as feminist metaethics.

Apart from articulating distinct questions that form its subject matter, feminist philosophy also tries to answer such questions in distinct, non-traditional ways. In like manner, feminist metaethics is distinguished not only by its key topics and their framing but also by its tools and methods. It draws together research strands which, despite their admittedly wide diversity, share an opposition to traditional metaethics. This opposition is expressed in the recognition that ethics should not be restricted to the analysis, synthesis and comparison of different theories but should likewise interrogate the ways in which different groups contribute to and are affected by ethical theories as well as the ways in which ethical curricula could engage or otherwise affect students [40]. To accomplish these aims, feminist metaethics disputes the assumption that 'those who do the theory know more about those who are theorized than vice versa' [45:575]. Hence, it discards ideal theory, which by relying 'on idealization to the exclusion, or at least marginalization, of the actual' ends up being 'in part ideological, in the pejorative sense of a set of group ideas that reflect, and contribute to perpetuating, illicit group privilege' [54:166-168]. In its place, it favours a theorisation of the non-ideal and takes a closer look at lived experience, history, (political) context and the perspectives of women as well as other oppressed groups. In short, what distinguishes feminist metaethics is a combination of two elements: its subject matter, as explained above, and, most importantly, its 'explicit willingness to engage in pedagogical self-examination and to critically consider one's own (historic, social) position, power and influence, or lack thereof' [40].

## 3 How Would a Feminist Metaethics of AI Look Like?

As emphasised in section 1, research on AI ethics abounds. Yet, despite the recent incidents giving AI ethics a bad name, to date such research has tended to focus on first-order AI ethics questions, that is, on normative and applied ethics, rather than second-order ones, that is, on metaethics. This gap in second-order reflection about AI ethics could be fruitfully mitigated by adopting the lens of feminist metaethics, thanks to its critical, normative nature.

In trying to sketch a preliminary research agenda for a feminist metaethics of AI, I provide an overview of the main directions that it could take. More concretely, a feminist metaethics of AI would examine: first, the continuity between AI ethics theory and action; second, the actual effects of the AI ethics discourse as opposed to the merely speculative ones; third, the agents involved in AI

---

[1] Jongepier uses 'pedagogical metaethics' and 'feminist metaethics' interchangeably [40]. Alternatively, 'critical metaethics' or 'radical metaethics' could be explored as related terms, but their scope would be different from that of the 'feminist metaethics' discussed here, since, notwithstanding their points of convergence and potential synergies, feminist theory and critical theory remain rooted in distinct traditions.

ethics; and finally, the choice of topics in AI ethics as well as key methods to approach said topics.[2]

## 3.1 Theory vs Action in AI Ethics

Ethics is mostly identified with ethical theory, but a feminist metaethics would not take for granted that ethical theory is needed in all cases. Considering that the numbers of professional philosophers keep growing, Baier [6] encourages us to question and demand a justification for why society should be supporting the activity of ethical theorising. Before philosophers attempt to advise the members of other professions about their ethics, they should turn their attention inwards and critically examine their own profession. In this direction, Baier [6:240] offers them a series of questions to guide their thinking: 'first, why anyone should be a philosopher (in our sense); second, why anyone else should pay one to be a philosopher; third, how many should be paid philosophers; and fourth, what exact form the professional activity of paid philosophers should take, and in particular what sort of examination of moral practices the moral philosopher should conduct'.

Accordingly, before delving into competing theories, principles and values to govern AI, AI ethicists should ask: do we even need AI ethics at all? Is AI really such a unique technology that merits its own research field, as the idea behind AI exceptionalism goes, or is philosophy falling for nothing more than a hype? Why cannot AI ethics be modelled upon, e.g., medical ethics [55]? Is AI ethics an activity that should be publicly funded? If yes, to what extent and by whom? Is it acceptable for technological companies or the military to fund research into the ethical implications of their core activities? Relatedly, should public-private partnerships on AI ethics be encouraged?

Another relevant question for a feminist metaethics of AI would be the relation between theory and action. In the words of MacIntyre [47:58], '[t]here ought not to be two histories, one of political and moral action and one of political and moral theorizing, because there were not two pasts, one populated only by actions, the other only by theories. Every action is the bearer and expression of more or less theory-laden beliefs and concepts, every piece of theorizing is a political and moral action.' When it comes to metaethics, as was previously mentioned, those espousing internalism support that one's moral judgements and one's behaviour are necessarily connected. Similarly, feminist metaethics is concerned with individuals' motivation to act upon the moral reasons that they acknowledge [78].

Hence, it could be examined whether AI ethics should translate to a commitment to action on a social level, e.g., in the form of political reform and solidarity or even law-making. The same question could be applied on a personal/individual level, e.g., questioning whether AI ethicists are or should be actually living by the principles or values that they are advancing. Should we assume that those trained in ethics are better people or that we can look up to them as exemplars [40]? Should the moral improvement of students be an explicit aim of training in AI ethics? The recent allegations against Joichi Ito, for example, cause at least some uneasiness about AI ethicists whose own practices do not abide by ethics standards. Along the same lines, it is worth wondering whether AI ethics should translate to action on a technological level. Should we operationalise AI ethics through ethics-by-design or FAT-ML (Fairness, Accountability, and Transparency in Machine Learning) techniques? On the one side, ethicists working on such techniques have the ability to collaborate closely with technologists and influence the development of a technological product or service from scratch. On the other side, doing so might constitute an instance of the 'mathematization of ethics' [10] or 'technological solutionism' [59], in the sense of adopting computational fixes for what are essentially social and political issues. Alternatively, Morley, Elhalal et al. [57] have advocated for a combination of ethical boards, codes, audits and tools termed 'Ethics as a Service'.

## 3.2 The Real-Life Effects of AI Ethics

For feminist metaethics, ethics is not restricted to an intellectual exercise; it has practical significance and social impact. Formulating abstract ethical theories or principles and seeking to apply them to concrete cases might have consequences that are, in the best scenario, unexpected and messy or, in the worst scenario, harmful. In what follows, I describe four adverse effects that the AI ethics discourse can have: ethics-washing, ethics-shopping, ethics as branding and ethics-bashing.

Ethics-washing denotes using the language of ethics for the sake of giving merely the appearance of ethical behaviour, while what is truly intended is to justify de-regulation or self-regulation [11]. A recurrent theme in the AI ethics literature is that regulation through legal instruments faces a pacing problem, which means that the law is too rigid and always lagging behind fast-paced technological development [50]. Equally, it is often feared that regulation through laws will stifle innovation and impair entire technological sectors [23]. In light of these shortcomings, ethics is presented as a sufficient, far more flexible and thereby preferable option than law.

Yet, this uncontested preference for ethics over law poses procedural problems and generates tension in the relation between ethics and law in the context of AI [83]. Compared to law-making, ethical decision-making is not equipped with democratic legitimacy, checks and balances, hard enforcement capabilities, or accountability and redress mechanisms. A technological company's compliance with its ethical guidelines or code of conduct rests solely upon its voluntary commitment, whereas possible violations of said guidelines and codes incur mostly reputational risks than tangible financial or legal penalties. This is why ethics recommendations and guidelines are often mentioned as 'soft law', or 'soft law governance of AI' in our specific context [49], compared to the 'hard law', which comprises binding legal instruments. For example, the aforementioned AI HLEG issued its Ethics Guidelines for Trustworthy AI which were fed directly into the policymaking process, yet it comprised only self-nominated experts, not any elected representatives [33]. Furthermore, it comprised far more industry actors than academics or civil society representatives, since almost half the group represented industry interests [41]. Not only that, but it is hard to see why good ethical reasoning and decision-making should be assumed to be faster than legislative procedures; after all, philosophers debate certain ethical issues for decades without reaching definitive conclusions.

---

[2] The order in which these directions are mentioned does not reflect any judgement on their significance or priority.

Ethics-shopping is defined as the 'malpractice of choosing, adapting, or revising ("mixing and matching") ethical principles, guidelines, codes, frameworks, or other similar standards [...] from a variety of available offers, in order to retrofit some pre-existing behaviours (choices, processes, strategies, etc.), and hence justify them a posteriori, instead of implementing or improving new behaviours by benchmarking them against public, ethical standards' [24:186]. The concern about ethics-shopping is likewise echoed in Baier [6], who laments the emergence of a marketplace of available theories and principles from which the professional philosopher picks and chooses based on their clients' needs.

Ethics as branding refers to the instrumentalisation of ethics as a means to promote business uptake. For instance, the AI HLEG and the European Commission speak of Ethical AI made in Europe. In one of its Communications, the latter says: 'Ethical AI is a win-win proposition. Guaranteeing the respect for fundamental values and rights is not only essential in itself, it also facilitates acceptance by the public and increases the competitive advantage of European AI companies by establishing a brand of humancentric, trustworthy AI known for ethical and secure products' [18:8]. Therefore, ethics acts as a seal of approval for AI-based products. As such, it serves the purpose of appeasing public criticism and pressure as well as helping re-brand European AI, without necessarily guaranteeing substantial ethical reforms within the industry.

These three effects are linked to a fourth and wider one. Although previously policymakers and civil society were concerned about technological companies not taking ethics seriously enough, there is now a risk of distrust or avoidance of invoking ethics when it comes to AI. Bietti [11] refers to this risk as 'ethics-bashing', which she explicates as the 'tendency [...] to trivialize "ethics" and "moral philosophy" by reducing more capacious forms of moral inquiry to the narrow conventional heuristics or misused corporate language they seek to criticize.'

The preceding effects lead one to wonder whether the AI ethics discourse does more harm than good in real life, as opposed to in theory, a question that would be crucial for a feminist metaethics of AI. Should we continue having AI ethics debates despite their potential side effects? If yes, how, when and where should such debates take place? Might our focus on these debates distract us from other, more important issues? Indicatively, it could be argued that the frequently witnessed practice of invoking thought experiments in the context of AI ethics, such as MIT's Moral Machine experiment [5], detracts from an exploration of its actual effects. Conversely, to help anticipate the real-life effects of AI ethics, we could avail of Baier's [6] guiding question: will the followers of x moral code become exploiters or exploited? More concretely, we could combine normative with empirical research, such as the survey by de Laat [42] on the implementation of ethics principles by AI companies, the research by McNamara et al. [52] on how codes of ethics influence developers' decision-making, or the most recent research by Morley, Kinsey, et al. [58] on AI practitioners' perspectives about ethics.

Also in accordance with the previously examined continuity between theory and action, a feminist metaethics of AI would not be restricted to a theoretical analysis of such effects but should expand to criticise malpractice and advocate for correction in favour of the marginalised. Phenomena such as ethics-washing, ethics-shopping, the use of ethics as a sales pitch and ethics-bashing deflect the focus of the conversation from protecting marginalised groups. Furthermore, they turn the ethical considerations surrounding AI into non-genuine, compliance procedures to make sure that the next AI product will tick all the necessary boxes before its development and deployment. Thus, in the context of AI, a feminist metaethics approach should pave the way towards enabling the detection, criticism and correction of such phenomena.

### 3.3 The Agents of AI Ethics

The growth observed in the AI ethics literature is accompanied by the rise of a new profession, that of the AI Ethicist. Management consulting firms stress that companies should include AI ethicists, Chief AI Ethics Officers and AI Ethics Councils in their composition and urge them to create dedicated positions for a single or, even better, multiple postholders [4, 21]. However, the job description of such a role and, subsequently, its candidates' eligibility requirements are far from clear. More broadly, the next question that a feminist metaethics of AI should handle is who should be doing AI ethics. In this regard, it is worth examining what a division of labour would be like in terms of the demarcation between theory and action but also in terms of the different disciplines engaged in AI ethics, i.e. law, philosophy, computer science and engineering, and business administration.

In relation to the 'divorce' of theory from action, Baier [6] observes the emergence of two types of agents doing ethics. On the one hand, there are the 'theorists', the 'beautiful souls', who confined in their ivory towers manage to dissociate themselves from the real-world implications of their theories. On the other hand, there are the applied ethicists. This 'new breed of professionals' acts as the conscience of other professions: their role is to gather all available ethical theories and, after gauging each one's costs and benefits, find the one that will best suit the needs of their clients, including decision-makers [6]. Highly critical of this division, Baier [6] follows Hume in suggesting that philosophers should, instead, embrace 'a share of the gross earthly mixture' that characterises non-philosophers and conduct their philosophical endeavours in ways that are empirically informed and non-intellectualist. By resisting armchair speculation and observing non-philosophers' work, they are likely to yield more accurate evaluations of the procedures involved [6]. Consequently, a feminist metaethics of AI would benefit from the insights of philosophers who have gained practical knowledge and first-hand experience on the subject matter they seek to evaluate, namely AI development.

With respect to the different disciplines, Jongepier suggests that, since everyone is capable of moral deliberation, we should entertain the thought that scholars who are not trained in ethics might be better at adjudicating ethical concerns [40]. First of all, it is not uncommon for law and ethics to tackle the same topics, often in the form of inter- and intra-disciplinary research. Especially when it comes to the governance of emerging technologies, research witnesses a growing 'ELSI-fication', in the sense that the Ethical, Legal and Social Implications (ELSI) of technology are dealt with in tandem [27]. However, on the one hand, legal scholars have tried to formulate strict conditions for the integration of ethical enquiry into their research [79]. On the other hand, philosophers such as Leiter [43] have argued against legal scholars invoking another discipline in ways that do not abide by such discipline's standards and have even accused them of 'intellectual voyeurism'. At the same time, although other types of advisors usually have some experience-based expertise, philosophical ones have theories that are only tested by argument. Are, then, maybe computer scientists better placed to answer ethical questions about AI, given that they possess such experience-based expertise? Or, could the same hold true for those with expertise in business administration and management? Google's recent dispute with Gebru is a case in point for examining if AI ethics can take place in a corporate context in light of potential conflicts of interest and, if so, which are the limitations of such a scenario, particularly for researchers' integrity and freedom of research, as well the safeguards needed to protect AI ethics researchers (e.g., arguably some sort of whistle-blower protection). On the same note, Abdalla and Abdalla have fruitfully compared the impact of Big Tech funding on AI ethics research with that of funding by the Big Tobacco industry on health-related research [1]. Last but not least, what about lay people? Are they not optimally placed to partake in ethical decision-making about the technologies that are or will be directly affecting them? Or, would collaboration among all these professionals and non-professionals be the suggested approach? More generally, what does expertise in AI ethics mean? These are questions with which a feminist metaethics of AI should grapple.

In any case, a feminist metaethics of AI would acknowledge that AI ethics is not made in a vacuum. It would urge all individuals and organisations involved to be critically aware of their situatedness and their own role in legitimising unequal relationships and situations. D'Ignazio and Klein [15] describe what they call the 'privilege hazard': 'the phenomenon that makes those who occupy the most privileged positions among us—those with good educations, respected credentials, and professional accolades—so poorly equipped to recognize instances of oppression in the world.' In the same vein, Baier [6:238] regrets that ethicists have not been 'very conspicuous for their sense of social answerability' and exhorts them to be more 'self-conscious about their own social role'. Likewise, we should keep in mind the questions that Jasanoff and Hurlbut [36:437] have put forward with respect to ethical debates, i.e., 'who sits at the table, what questions and concerns are sidelined, and what power asymmetries are shaping the terms of debate'. Thus, the agenda for a feminist metaethics of AI would, first, include research into the composition of AI ethics boards and committees in terms of their diversity and representativeness and, second, demand that those hitherto ignored are given more opportunities to join decision-making processes.

### 3.4 Topics and Methods in AI Ethics

What should be the subject of ethical enquiry admits different interpretations. As Bartlett [7:837] notes '[a] question becomes a method when it is regularly asked. Feminists across many disciplines regularly ask [...] "the woman question," [...] to identify the gender implications of rules and practices which might otherwise appear to be neutral or objective'. Likewise, feminist metaethics focuses its theoretical and practical enquiry on specific topics, particularly—yet not exclusively—on those relevant to women's lives and experiences. A feminist metaethics of AI would examine the impact of ostensibly neutral AI systems albeit not on people in general. Conceding that 'no one is simply a person but instead is constituted fundamentally by race, class, and gender' [35:163], it would examine said impact depending on one's gender as well as, through an intersectional feminist lens, depending on other identity markers with which gender interacts and which affect one's privilege or disadvantage, including class, race, disability, sexuality, and so on. Who is being left out from such an examination is also important, given that, for some feminist scholars, moral ignorance does not always exculpate [56].

Contrary to mainstream ethics, which, following ideal theory approaches, tends to 'abstract away from relations of structural domination, exploitation, coercion, and oppression' [54:168], a feminist metaethics of AI would make such relations visible and assess how these are shifted by AI. Indicatively, an interrogation of power relations might include the power dynamics and diversity within the AI workforce itself, examining who has the computational power and resources to actually train AI systems and benefit from them, or concerns of data colonialism, which Couldry and Mejias [12] describe as 'the appropriation of human life so that data can be continuously extracted from it for profit'. Equally, it is worth researching how AI systems affect power relations in terms of who is believed or not as well as if and how they might exacerbate testimonial injustice [25]. In addition, as Superson [78] notes, feminists are often doubtful of moral absolutism, which acknowledges just one true moral code, as they are worried about judging other cultures and eventually slipping into moral imperialism. Hence, power relations could also be investigated from the perspective of international relations. Such an investigation could include: the 'ghost work' behind the training of AI systems by click workers in the Global South [28, 34]; the practice of 'ethics-dumping', which occurs when the research and development of AI are carried out outside the place of origin and in ways that would be ethically unacceptable there [24]; the role of the EU as an 'exporter' of AI ethics, who seems to favour a one-way exchange of values as a prerequisite for collaboration rather than a bilateral one; and, finally, the dominance of economically developed countries and the underrepresentation of areas such as Africa, South and Central America, and Central Asia in the international AI ethics debate [37].

A feminist metaethics of AI would also use methods that deviate from those usually employed in ethics. Traditionally, ethics involves a rejection of the private and the situated over 'male-centred issues of the public and the abstract' [31]. By contrast, feminist ethics emphasises consideration of 'empirical information and material actualities' [65]. This is because by focusing on oppression, it tries to shed light on the hitherto ignored perspectives of those marginalised and oppressed. Hence, feminist metaethics entails a shift away from abstract, ideal cases and their allegedly essential features and towards a painstaking attentiveness to and engagement with actual experiences of individuals, their context and the relationships involved [74]. One way for AI ethics to encompass this careful attention would be by listening to and amplifying testimonies of situated agents. In doing so, a feminist metaethics of AI would take a narrative approach to ethics in the sense of attributing a normative role to the telling and hearing of stories. Specifically, stories can be invoked: '(1) to teach us our duties, (2) to guide morally good action, (3) to motivate morally good action, (4) to justify action on moral grounds, (5) to cultivate our moral sensibilities, (6) to enhance our moral perception, (7) to make actions of persons morally intelligible, and (8) to reinvent ourselves as better persons' [31]. Apart from the testimonies of physically, socially and historically situated agents, our moral imagination would be expanded by invoking stories from the realm of fiction. Particularly in the context of AI ethics, science fiction provides ample resources from which to draw.

Beauchamp [8] favours another similarly concrete way, which is more popular in law and business administration than ethics, that of case studies. He insists on a reversal of the most common top-to-bottom way of thinking, i.e., from high-level theory to applied contexts, as he believes that by inductively reflecting on applied contexts one can extract broader theories. Illustrative case studies in AI ethics are, for example, Google's image recognition algorithm that classified a black person as a gorilla [19], the AI gaydar developed by Stanford researchers [61], the gendered effects of AI automation, and the use of humanoid robots for care and sex work.

What is more, thinking about the practical significance of ethical theories implies not only an anticipation of their real-world effects, as described in subsection 3.2, but also changes in the ways we invoke philosophical concepts. As Mills [54:175] cautions, 'one has to be self-conscious about the concepts that "spontaneously" occur to one, since many of these concepts will not arise naturally but as the result of social structures and hegemonic ideational patterns'. When choosing between different concepts or schema, feminist metaethics would, first, consider the normative implications that such choices hold for the way we represent the world and, second, seek to adopt the concepts that are most ethically apt and best serve their practical purposes, such as social justice. It would ask: 'What is the point of having the concept in question? What concept (if any) would do the work best?' [32]. For instance, Tigard [82] underlines the potentially harmful effects of appealing to the concept of responsibility with regard to AI. As another example, the concepts of justice and autonomy are frequently invoked in AI ethics; yet, these concepts are contested and often criticised as reflecting masculine ideals. Therefore, in trying to choose between competing understandings, a feminist metaethics of AI would prioritise an intersectional conceptualisation of justice [13, 32] and a relational conceptualisation of autonomy [26, 48, 63], as these would better serve its empowerment purposes across different axes of oppression.

At the same time, philosophy has overemphasised the role of reason, neutrality and objectivity at the expense of emotions and affect, which are traditionally associated with women or femininity and are viewed with hostility as being irrelevant or even a detriment to moral judgement. However, for feminism, 'affect is necessary to knowing the moral landscape, making correct moral judgments, and acquiring moral truths' [78]. This is because emotions, and especially caring about something or someone, foster one's awareness of and receptivity to the particularities of the situation at hand [78]. Thus, *contra* Kant's pure practical reason and the 'myth of the dispassionate investigator', a feminist metaethics of AI would acknowledge a 'feedback loop between our emotional constitution and our theorising' [35:164-170]. Consequently, it would accommodate examinations of (intellectual but not only) emotions, such as 'loving perception' or 'sympathetic thinking' [44], 'philosophical love' [39], or even anger [9, 75]. It would also take into account the relationships and dependencies of moral agents, adopting the methods of 'care ethics' [70]. As a general rule, without an affective engagement with the issue and the persons involved, AI ethics is not expected to yield accurate conclusions.

## 4 Where Does This Leave Us?

By and large, AI ethics proceeds from an interest in the competing values and principles that should guide the development and deployment of AI systems. However laudable, as a series of recent problematic incidents about AI ethics and AI ethicists demonstrate, this orientation is insufficient and might even backfire. Hence, this paper tries to draw attention to the significance of second-order reflection on AI ethics, suggest the intensification of existing relevant enquiries as well as their expansion into new topics and methods, and sketch a potential structure for a more systematic organisation of such a reflection. Echoing the proverbial camel that does not see its own hump,[3] if AI ethicists are to evaluate the conduct and work outputs of other professions, they must start by critically evaluating their own profession and avoid being blind to their own ethical pitfalls, such as the ones described at the beginning of this paper.

In particular, I sought to develop a roadmap for what a feminist metaethics of AI would like. Contrary to traditional metaethics, which involves reflecting on the nature of morality and moral judgements in a descriptive/non-normative way, feminist metaethics expands its scope to look beyond what morality actually is and into what our engagement with ethics should be like. This second-order reflection would benefit the sub-field of AI ethics. Such an application of feminist metaethics to the context of AI would give rise to a feminist metaethics of AI, for which I suggest four main lines of inquiry. Specifically, I argued that a feminist metaethics of AI would examine the assumed primacy of ethical theory over action, the real-life effects of AI ethics, the role and profile of those involved in AI ethics, topics related to power imbalances and inequalities, and methods that encompass attention to context, emotions and narrative.

---
[3] Proverb found in the Greek and Arabic tradition to denote those who see others' faults but not their own. The equivalent saying in English would be 'the pot calling the kettle black'.

The roadmap that I provided in this paper is admittedly broad and preliminary, and much research is needed to further flesh out its details and boundaries. It does not purport to represent the sole or an exhaustive approach to feminist philosophy or feminist metaethics, but merely seeks to identify core elements that are shared by feminist approaches. Neither does this paper purport to put forward entirely new topics or ideas for discussion. Instead, building on existing literature, it makes a start on a large systematising effort. As a result, it brings together previously disparate criticisms raised by scholars of different backgrounds (e.g., law, ethics, science and technology studies, feminist theory) under the umbrella of a 'feminist metaethics of AI' agenda.

Overall, the way forward for an AI ethics informed by feminism involves paying attention to the complexity of the moral landscape and advocating for an enlargement of what is deemed as the orthodox scope of metaethics. On a theoretical level, this endeavour could offer compelling insights for future research on AI ethics. As a welcome corollary, it might have pedagogical value, converting otherwise different critiques into a body of thought accessible to newcomers in the AI ethics field. On a practical level, it could serve as a useful resource for policymakers, legislators and practitioners working on AI and its governance, helping them steer away from the misuse of ethics. More broadly, though, it aspires to move critical conversations about the role of AI ethics beyond the ivory tower and into the real world.

## ACKNOWLEDGMENTS

The author's research is funded by the Fonds Wetenschappelijk Onderzoek (FWO, Research Foundation – Flanders) as part of a PhD Fellowship for fundamental research (no. 1151621N). Special thanks are owed to the participants of the 2021 OZSW Spring School 'Moral Theory and Real Life' and particularly to its organisers, Dr Sven Nyholm and Dr Fleur Jongepier, for valuable feedback on an earlier version of this paper. The author would also like to thank the anonymous AIES reviewers for their insightful comments, which helped improve the manuscript.